\documentclass[runningheads]{llncs}
\usepackage[utf8]{inputenc}
\usepackage[T1]{fontenc}
\usepackage{hyperref,bookmark}
\usepackage{booktabs}
\usepackage{listings,chngcntr}
\AtBeginDocument{\counterwithin{lstlisting}{section}}
\usepackage{todonotes}
\usepackage{amsmath}
\usepackage{amssymb}
\usepackage{algorithm,algpseudocode}
\usepackage{multicol}
\usepackage{comment}
\usepackage{tikz}
\usetikzlibrary{shapes,positioning,arrows,fit,trees,calc}

\newcommand{\etal}{et al.}

\begin{document}

\title{JEFL: Joint Embedding of Formal Proof
  Libraries}
\titlerunning{Joint Embedding of Formal Proof Libraries}

\author{Qingxiang Wang\inst{1}
\and Cezary Kaliszyk\inst{1,2}\orcidID{0000-0002-8273-6059}}
\authorrunning{Q. Wang \and C. Kaliszyk}

\institute{University of Innsbruck, Austria
\and University of Warsaw, Poland
\email{shawn.wangqingxiang@gmail.com,cezary.kaliszyk@uibk.ac.at}}

\maketitle

\begin{abstract}
The heterogeneous nature of the logical foundations used in different interactive proof assistant libraries has rendered discovery of similar mathematical concepts among them difficult.
In this paper, we compare a previously proposed algorithm for matching
concepts across libraries with our
unsupervised embedding approach that can help us retrieve similar concepts.
Our approach is based on the \texttt{fasttext} implementation of Word2Vec, on top of which a tree traversal module is added to adapt its algorithm to the representation format of our data export pipeline.
We compare the explainability, customizability, and online-servability of the approaches and argue that the neural embedding approach has more potential to be integrated into an interactive proof assistant.

\end{abstract}

\section{Introduction}\label{s:intro}

One of the
challenges hindering massive formalization of mathematics is the heterogeneous nature of the logical frameworks used in various interactive proof assistants \cite{harrison-2009,Paulson88,mizar-in-a-nutshell,HuetH14}.
When formalizing proofs against one formal library, it is informative to explore whether and how similar things are done in other libraries.
Such exploration has to be done manually and would usually require expertise in the other proof assistants.
It would be nice if a tool could
let users more systematically explore and discover commonality among formal libraries.

Not only can such a tool be an informative recommender when integrated into an interactive proof assistant, but exploring commonalities among formal libraries is also an interesting problem per se.
Through time, multiple versions of the same or similar mathematical concepts have been formalized separately, resulting in repetitive work \cite{Klein14}.
To the mathematically oriented, it is quite irksome that identical mathematical concepts must require idiosyncratic formalizations in order 
to achieve assurance.
We believe that by investigating their commonalities, insights on improving interoperability among proof assistants can be obtained, thereby advancing the frontiers of combining systems.

Previous works~\cite{tgck-cicm14,tgck-jsc19} on this problem let us obtain a data export pipeline that could transform data from six proof assistants into a common term representation format (Fig.~\ref{fig:arch}),
on top of which an
iterative pattern-matching algorithm that could output constant/theorem pairs with high similarity scores was invented by Gauthier.
The alignments between the concepts across multiple proof libraries or within one library have been useful for tasks including
conjecturing \cite{tgckju-cicm16},
browsing multiple libraries simultaneously \cite{dmtgckmkfr-cicm17}, and
proof automation using learned alignments \cite{tgck-lpar15}.

The Gauthier approach, while being remarkably effective and useful, lacks \textit{explainability}, \textit{customizability} and \textit{online-servability} that hamper its integration into proof assistants.
By these three notions we mean the lack of \textit{mathematical intuitiveness}, lack of \textit{room for customization}, and lack of \textit{possibility for system integration}, respectively.
We introduce an alternative embedding approach based on the superb engineering of the \texttt{fasttext} implementation~\cite{bojanowski-etal-2017-enriching}.
This new approach could potentially overcome these drawbacks while providing competitive performance.
It could also serve as a highly configurable experiment platform for studying the alignment of multiple proof assistant libraries.
We coin this research \textit{JEFL}, as an acronym for \textit{Joint Embedding of Formal Libraries}.

\pgfdeclarelayer{background}
\pgfdeclarelayer{foreground}
\pgfsetlayers{background,main,foreground}

\tikzstyle{export}=[draw, fill=blue!20, text width=5em, text centered, minimum height=2.5em, rounded corners]
\tikzstyle{gauthier}=[draw, fill=red!20, text width=6em, text centered, minimum height=2.5em, rounded corners]
\tikzstyle{jefl}=[draw, fill=green!20, text width=3.5em, text centered, minimum height=2.5em, rounded corners]
\tikzstyle{to}=[->,>=stealth',shorten >=2pt,thick,font=\sffamily\footnotesize]
\def\blockdist{3.0}
\def\blockdistnarrow{1.6}

\begin{figure}[tb]
\begin{center}
\begin{tikzpicture}
\node (hol4) [export] {HOL4};
\node (hollight) [export,below of=hol4] {HOL-Light};
\node (isabelle) [export,below of=hollight] {Isabelle};
\node (coq) [export,below of=isabelle] {Coq};
\node (matita) [export,below of=coq] {Matita};
\node (mizar) [export,below of=matita] {Mizar};
\node (exportlabel) [below of=mizar] {Exporters};

\path (coq)+(\blockdist,-0.75) node (preprocess) [gauthier] {Preprocess};
\node (pattern) [gauthier,above of=preprocess] {Patternify};
\node (score) [gauthier,above of=pattern] {Scoring Iteration};
\node (gaumain) [gauthier,above of=score] {Main};
\path (mizar)+(\blockdist,0) node (io) [gauthier] {IO, Parsing...};
\node (gaulabel) [below of=io] {Gauthier \etal};

\path (io)+(\blockdist,0) node (tt) [jefl,fill=purple!20] {\texttt{tt}};
\path (tt)+(\blockdistnarrow,0) node (vector) [jefl] {Vector};
\path (vector)+(\blockdistnarrow,0) node (matrix) [jefl] {Matrix};
\node (utils) [jefl,above of=tt] {Utils};
\node (args) [jefl,above of=vector] {Args};
\node (dict) [jefl,above of=matrix] {Dict};
\node (model) [jefl,above of=args,text width=10em] {Model};
\node (fasttext) [jefl,above of=model,text width=10em] {FastText};
\node (jeflmain) [jefl,above of=fasttext,text width=10em] {Main / Server};
\node (jefllabel) [below of=vector] {JEFL};

\begin{pgfonlayer}{background}
\path (hol4.west |- hol4.north)+(-0.2,0.2) node (a) {};
\path (exportlabel.south -| exportlabel.east)+(+0.4,-0.2) node (b) {};
\path [fill=blue!10,rounded corners, draw=black!50, dashed] (a) rectangle (b);

\path (gaumain.west |- gaumain.north)+(-0.3,0.2) node (c) {};
\path (gaulabel.south -| gaulabel.east)+(+0.2,-0.2) node (d) {};
\path[fill=red!10,rounded corners, draw=black!50, dashed] (c) rectangle (d);

\path (utils.west |- jeflmain.north)+(-0.3,0.2) node (e) {};
\path (jefllabel.south -| matrix.east)+(+0.2,-0.2) node (f) {};
\path[fill=green!5,rounded corners, draw=black!50, dashed] (e) rectangle (f);
\end{pgfonlayer}

\begin{pgfonlayer}{foreground}
\draw[to] (b |- io) -- (io) node[midway,above] {\texttt{tt}};
\draw[to] (io) -- (preprocess);
\draw[to] (io) -- (e |- io) node[midway,above] {\texttt{sexp}};
\end{pgfonlayer}

\end{tikzpicture}
\end{center}
\caption{The architectural relationship between Gauthier's approach and JEFL. At the current stage, the exporters dump text in the \texttt{tt} format (Section~\ref{s:prev}). JEFL reuses the IO/parsing module of Gauthier and passes s-expressions to the \texttt{fasttext} implementation.}
\label{fig:arch}
\end{figure}
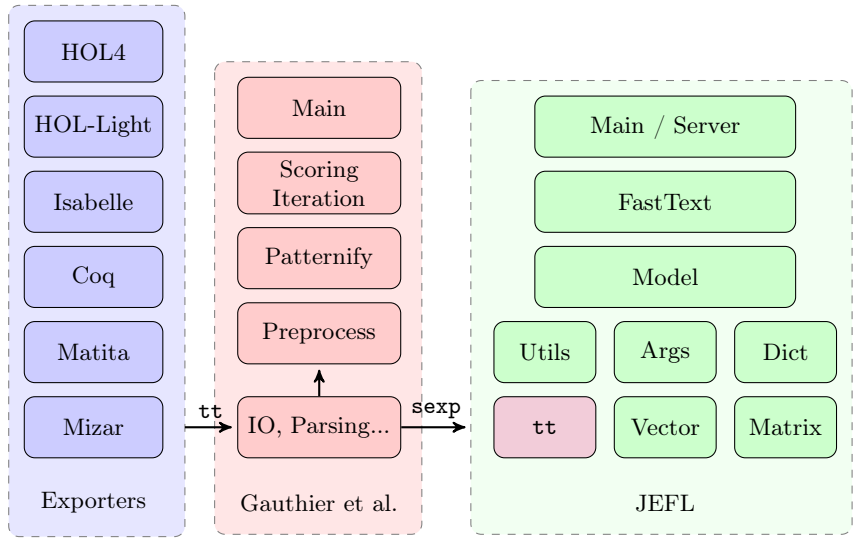

\section{Previous Works and the \texttt{tt} Format}\label{s:prev}

Exchanging formal developments within or across formal systems has been studied through three strands of research.
First, on the library translation side, many tools that can partially translate proofs have been developed, including those from HOL to Isabelle/HOL \cite{obua-skalberg-2006}, HOL Light to Coq \cite{keller-werner-2010}, HOL Light to Isabelle/HOL \cite{KaliszykK13}, respectively.
Second, on the ontology sharing side, Bortin~\cite{bortin-2006}, Rabe~\cite{rabe-cicm2013}, Hurd~\cite{hurd-opentheory-2011}, So and Watt~\cite{so-mathml2openmath}, and Carlisle~\etal~\cite{carlisle-mathml2openmath} each made their own contribution translating specifications or formal proof objects between formal or semi-formal mathematical representations.
These two strands of research mostly either solely provide guidelines on manual processing or require manual work at a certain phase of their framework.

The third strand comes from enhancements of ITP systems.
Heras and Komendantskaya~\cite{keras-komendantskaya} implemented a recurrent term clustering algorithm to find proof similarities in Coq/SSReflect libraries.
Urban~\cite{Urban06-ijait} created tools for large-scale retrieval of the Mizar Mathematical Library into a clausal format.
Kaliszyk and Urban~\cite{KaliszykU13a} exported the core HOL Light library as well as the Flyspeck~\cite{Hales05} library to evaluate the relevance of lemmas by combining the power of automated theorem provers.
This work was later extended to a web service \cite{hhmcs} and experimented with using multiple representation formats and different automated theorem provers in~\cite{holyhammer}.

A byproduct along~\cite{KaliszykU13a,hhmcs,holyhammer} was a collection of exporting and post-processing techniques specific to HOL Light, including a TPTP-style~\cite{Sutcliffe10} data representation format which we internally called ``the \texttt{tt} format''.
The formalism of \texttt{tt} is based on a simple term structure that is flexible enough to represent the kernel representations of formal data on diverse logical foundations, so there is a potential to export data from multiple proof assistants into this common format.
Based on the export of three HOL libraries (HOL Light, HOL4, and Isabelle/HOL), Gauthier and Kaliszyk proposed the first version of their scoring algorithm~\cite{tgck-cicm14} and used
for various conjecturing and transfer learning tasks.
A more comprehensive set of alignment experiments refined the scoring algorithm, provided a more uniform pattern-matching and guaranteed convergence, and was used on six proof assistant libraries (adding Coq, Matita, and Mizar)~\cite{tgck-jsc19}.

Listing~\ref{lst:21} and~\ref{lst:22} illustrate the definition of the predecessor of the naturals (\texttt{PRE}) of HOL Light being translated into a list of three \texttt{tt} items.
The last arguments of them can be parsed into term structures (Fig.~\ref{fig:ttterm}) using the type definition in Listing~\ref{lst:22}.

\begin{lstlisting}[language=bash,basicstyle=\ttfamily\scriptsize,caption={Definition of the predecessor of the naturals \texttt{PRE} in HOL Light.},captionpos=top,label={lst:21}]
-----------------------------------------------------------------------------
let PRE = new_recursive_definition num_RECURSION
 `(PRE 0 = 0) /\
  (!n. PRE (SUC n) = n)`;;
-----------------------------------------------------------------------------
\end{lstlisting}

\begin{lstlisting}[language=bash,basicstyle=\ttfamily\scriptsize,caption={\texttt{PRE} transformed to three \texttt{tt} items.},captionpos=top,label={lst:22}]
-----------------------------------------------------------------------------
01. tt('const/arith/PRE', ty, ('type/nums/num' > 'type/nums/num')).
02. tt('thm/arith/PRE_0', ax,
    (('const/arith/PRE' ('const/nums/NUMERAL' 'const/nums/_0')) =
      ('const/nums/NUMERAL' 'const/nums/_0'))).
03. tt('thm/arith/PRE_1', ax, 
    (![n : 'type/nums/num']:
      (('const/arith/PRE' ('const/nums/SUC' n)) = n))).
-----------------------------------------------------------------------------
\end{lstlisting}

\begin{lstlisting}[language=bash,basicstyle=\ttfamily\scriptsize,caption={Type definition of \texttt{tt} term in OCaml for parsing.},captionpos=top,label={lst:23}]
-----------------------------------------------------------------------------
type ttterm =
| Id of string (* may be a constant or variable *)
| Comb of ttterm * ttterm
| Abs of string * ttterm * ttterm;;
-----------------------------------------------------------------------------
\end{lstlisting}

The HOL Light and HOL4 exports directly use HOLyHammer's export \cite{holyhammer}.
For Isabelle, an ML component was implemented that extracts all theorems of the theory and writes them together with the declared constants and types in a text file.
The Coq export to the \texttt{tt} format was implemented by Gauthier as part of his work~\cite{tgck-jsc19}.
For Mizar, we rely on Urban's MPTP pipeline \cite{urban-mptp02} and transform the intermediate XML2 representation.

\tikzstyle{term}=[text width=1em, text centered, font=\ttfamily\scriptsize]
\tikzstyle{label}=[xshift=1mm, yshift=4mm,font=\ttfamily\scriptsize]

\begin{figure}[h!]
\begin{tikzpicture}[level distance=8mm, sibling distance=10mm]
\node(a01)[term]{Comb}
 child {
  node(a02)[term]{Id(:)}
 }
 child {
  node(a03)[term]{Comb}
  child { node(a04)[term]{Id(PRE)} }
  child {
   node(a05)[term]{Comb}
   child { node(a06)[term]{Id(>)} }
   child {
    node(a07)[term]{Comb}
    child { node(a08)[term]{Id(num)} }
    child { node(a09)[term]{Id(num)} }
   }
  }
 };
\node at([label]a01) {01};
\node(b01)[term,right of=a01,xshift=32mm,yshift=12mm]{Comb}
 child { node(b02)[term]{Id(=)} }
 child {
  node(b03)[term]{Comb}
  child {
   node(b04)[term]{Comb}
   child { node(b06)[term]{Id(PRE)} }
   child {
    node(b07)[term]{Comb}
    child { node(b10)[term]{Id(NUMERAL)} }
    child[missing]
    child { node(b11)[term]{Id(0)} }
    }
  }
  child[missing]
  child[missing]
  child {
   node(b05)[term]{Comb}
   child { node(b06)[term]{Id(NUMERAL)} }
   child[missing]
   child { node(b06)[term]{Id(0)} }
  }
 };
\node at([label]b01) {02};
\node(c01)[term,right of=b01,xshift=27mm,yshift=5mm]{Comb}
 child { node(c02)[term]{Id(!)} }
 child {
  node(c03)[term]{Abs}
  child { node(c04)[term]{n} }
  child { node(c05)[term]{Id(num)} }
  child {
   node(c06)[term]{Comb}
   child { node(c07)[term]{Id(=)} }
   child {
    node(c08)[term]{Comb}
    child {
     node(c09)[term]{Comb}
     child { node(c11)[term]{Id(PRE)} }
     child[missing]
     child {
      node(c12)[term]{Comb}
      child { node(c13)[term]{Id(SUC)} }
      child[missing]
      child { node(c14)[term]{Id(n)} }
     }
    }
    child[missing]
    child { node(c10)[term]{Id(n)} }
   }
  }
 };
\node at([label]c01) {03};
\end{tikzpicture}
\caption{Term structures of the definition of \texttt{PRE}. The tokens \texttt{PRE}, \texttt{num}, \texttt{NUMERAL} and \texttt{SUC} are short for \texttt{const/arith/PRE}, \texttt{type/nums/num}, \texttt{const/nums/NUMERAL} and \texttt{const/nums/SUC}, respectively. Note that the constant \texttt{const/arith/PRE} is included into the term \textit{01} with a type assignment operator \texttt{':'}. This allows embedding vectors to be assigned to the definition constants.}
\label{fig:ttterm}
\end{figure}
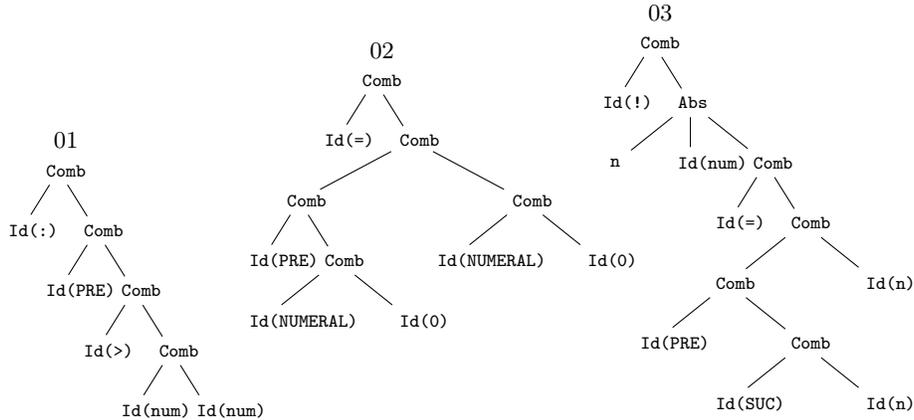

\section{The Architecture of JEFL}\label{s:arch}

In this paper, we
focus on the core similarity discovery algorithm.
Our claim on advantages in JEFL is with respect to the algorithmic part of the system.
We leave the eventual integration of the whole framework into proof assistants, with issues such as handling constants that have not been encountered during training, as future work.

\subsection{Similarity through Embedding}\label{ss:embedding}

A natural way to find similarities among concepts is to treat our problem as a distributed representation learning task.
Generally speaking, given a structure composed of atomic units, distributed representation learning seeks to represent each of the atomic units with a low-dimensional vector.
In effect, all the units are embedded into a Euclidean space, with their coordinates respecting the overall structure.
The notion \textit{distributedness} comes from the fact that the vocabulary size of a corpus is much larger than the dimension of a vector, and the information of an atomic unit is distributed in the coordinates of a vector.

The vectors are learned by analyzing the \textit{context} of each unit,~i.e. the information of units adjacent to or surrounding a target unit.
Once vector representations for units are learned, similarity between units can then be computed by cosine similarity with a range from $[-1,1]$.
For a set of units, vector representation can be computed by taking average of the vectors, and then similarity between different sets of units can also be computed using cosine similarity.

Notable unsupervised distributed representation learning algorithms include Pennington~\etal's GloVe algorithm~\cite{pennington2014glove} and Mikolov~\etal's Word2Vec algorithm~\cite{word2vec-a,word2vec-b}.
In this paper, we use Mikolov's Word2Vec algorithm.
Word2Vec works on texts or lists of word tokens.
For each word in the training corpus, a randomized span of words surrounding that word is picked to form the context of that word.
The context is then consumed by the Word2Vec model to conduct one step of the stochastic gradient descent updates.

\subsection{Adaptation of the \texttt{tt} Format in Word2Vec}\label{ss:adapt}

To illuminate our technique, it is interesting to note that DeepWalk~\cite{deepwalk} and Node2Vec~\cite{node2vec}, two methods on embedding large networks, also use Word2Vec as their underlying algorithm.
The data used by DeepWalk and Node2Vec are single-graph datasets with nodes that contain heterogenous information such as social profile details.
To fit Word2Vec, first the node information of the graph has to be transformed into a dictionary through data processing.
Then we perform random walks along the paths of the graph to generate node sequences that resemble text corpus.
For each node in a node sequence, the corresponding context is generated as a span of nodes surrounding that node.

In our case, different from DeepWalk and Node2Vec, the formal library data in the \texttt{tt} format are not a single graph but a collection of trees.
More precisely, in order to compare two libraries, we need two lists of \texttt{tt} items from the two libraries to provide as training data.
The \texttt{tt} items are parsed as trees and then traversed in different ways to create sequences of node constants.
Examples of traversals include preorder, inorder, postorder traversals and their reverses, random walks from the root to a leaf, or just dump the leaves of a tree in some order.
With clever design, these traversals can also be combined to create hybrid orders.
At the current phase we implemented a simple weighted mechanism combining preorder, inorder and postorder traversals.
The weights of traversals are used to control the learning rate for SGD updates and are hyperparameters determined before training (Fig.~\ref{fig:traverse}).
We anticipate further experimental insights when other forms of traversals are implemented in the future.

\tikzstyle{term}=[text width=1em,text centered,font=\ttfamily\scriptsize]
\tikzstyle{tflag}=[text width=0.2em,text centered,font=\ttfamily\tiny,inner sep=0,outer sep=0]
\tikzstyle{arro}=[->,>=stealth',shorten >=0.5pt,purple,rounded corners,line width=0.5pt]
\tikzstyle{preorder}=[draw,fill=blue!10,rounded rectangle,dashed,text width=24em,align=right,font=\ttfamily\scriptsize,inner sep=1.0mm,outer sep=1.0mm]
\tikzstyle{inorder}=[draw,fill=red!10,rounded rectangle,dashed,text width=14em,align=right,font=\ttfamily\scriptsize,inner sep=1.0mm,outer sep=1.0mm]
\tikzstyle{postorder}=[draw,fill=green!10,rounded rectangle,dashed,text width=14.9em,align=right,font=\ttfamily\scriptsize,inner sep=1.0mm,outer sep=1.0mm]
\newcommand*{\mybox}[1]{\framebox{\strut #1}}

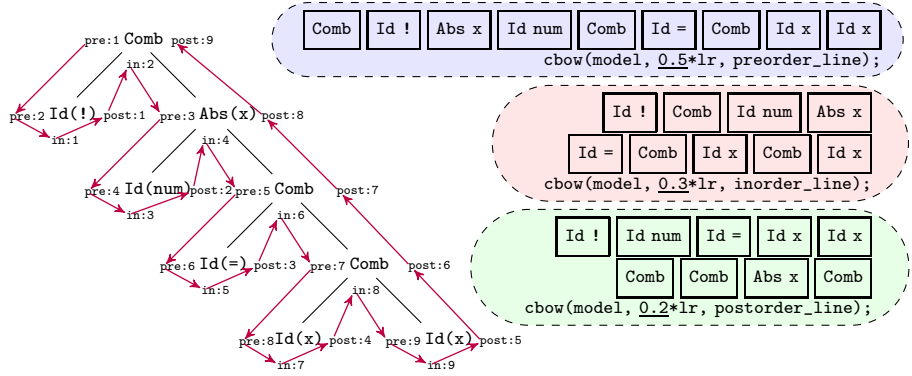
\begin{figure}[h!]
\begin{center}
\begin{tikzpicture}[level distance=10mm, sibling distance=20mm]
\node(x01)[term]{Comb}
 child { node(x02)[term]{Id(!)} }
 child {
  node(x03)[term]{Abs(x)}
  child { node(x04)[term]{Id(num)} }
  child {
   node(x05)[term]{Comb}
   child { node(x06)[term]{Id(=)} }
   child {
    node(x07)[term]{Comb}
    child { node(x08)[term]{Id(x)} }
    child { node(x09)[term]{Id(x)} }
   }
  }
 };
\draw[-] (x01) +(-0.65,-0.05) node (x01pre) [tflag] {pre:1}; 
\draw[-] (x01) +(-0.10,-0.35) node (x01in) [tflag] {in:2};
\draw[-] (x01) +(+0.50,-0.04) node (x01post) [tflag] {post:9};
\draw[-] (x02) +(-0.65,-0.05) node (x02pre) [tflag] {pre:2}; 
\draw[-] (x02) +(-0.10,-0.35) node (x02in) [tflag] {in:1};
\draw[-] (x02) +(+0.60,-0.04) node (x02post) [tflag] {post:1};
\draw[-] (x03) +(-0.65,-0.05) node (x03pre) [tflag] {pre:3}; 
\draw[-] (x03) +(-0.10,-0.35) node (x03in) [tflag] {in:4};
\draw[-] (x03) +(+0.70,-0.04) node (x03post) [tflag] {post:8};
\draw[-] (x04) +(-0.65,-0.05) node (x04pre) [tflag] {pre:4}; 
\draw[-] (x04) +(-0.10,-0.35) node (x04in) [tflag] {in:3};
\draw[-] (x04) +(+0.75,-0.04) node (x04post) [tflag] {post:2};
\draw[-] (x05) +(-0.65,-0.05) node (x05pre) [tflag] {pre:5}; 
\draw[-] (x05) +(-0.10,-0.35) node (x05in) [tflag] {in:6};
\draw[-] (x05) +(+0.70,-0.04) node (x05post) [tflag] {post:7};
\draw[-] (x06) +(-0.65,-0.05) node (x06pre) [tflag] {pre:6}; 
\draw[-] (x06) +(-0.10,-0.35) node (x06in) [tflag] {in:5};
\draw[-] (x06) +(+0.60,-0.04) node (x06post) [tflag] {post:3};
\draw[-] (x07) +(-0.65,-0.05) node (x07pre) [tflag] {pre:7}; 
\draw[-] (x07) +(-0.10,-0.35) node (x07in) [tflag] {in:8};
\draw[-] (x07) +(+0.65,-0.04) node (x07post) [tflag] {post:6};
\draw[-] (x08) +(-0.60,-0.05) node (x08pre) [tflag] {pre:8}; 
\draw[-] (x08) +(-0.10,-0.35) node (x08in) [tflag] {in:7};
\draw[-] (x08) +(+0.60,-0.04) node (x08post) [tflag] {post:4};
\draw[-] (x09) +(-0.65,-0.05) node (x09pre) [tflag] {pre:9}; 
\draw[-] (x09) +(-0.10,-0.35) node (x09in) [tflag] {in:9};
\draw[-] (x09) +(+0.60,-0.04) node (x09post) [tflag] {post:5};
\draw[arro] (x01pre) -- (x02pre);
\draw[arro] (x02pre) -- (x02in);
\draw[arro] (x02in) -- (x02post);
\draw[arro] (x02post) -- (x01in);
\draw[arro] (x01in) -- (x03pre);
\draw[arro] (x03pre) -- (x04pre);
\draw[arro] (x04pre) -- (x04in);
\draw[arro] (x04in) -- (x04post);
\draw[arro] (x04post) -- (x03in);
\draw[arro] (x03in) -- (x05pre);
\draw[arro] (x05pre) -- (x06pre);
\draw[arro] (x06pre) -- (x06in);
\draw[arro] (x06in) -- (x06post);
\draw[arro] (x06post) -- (x05in);
\draw[arro] (x05in) -- (x07pre);
\draw[arro] (x07pre) -- (x08pre);
\draw[arro] (x08pre) -- (x08in);
\draw[arro] (x08in) -- (x08post);
\draw[arro] (x08post) -- (x07in);
\draw[arro] (x07in) -- (x09pre);
\draw[arro] (x09pre) -- (x09in);
\draw[arro] (x09in) -- (x09post);
\draw[arro] (x09post) -- (x07post);
\draw[arro] (x07post) -- (x05post);
\draw[arro] (x05post) -- (x03post);
\draw[arro] (x03post) -- (x01post);

\node(preorder)[preorder,left of=x01,xshift=+70mm,yshift=-0.2mm]{\mybox{Comb} \mybox{Id !} \mybox{Abs x}  \mybox{Id num} \mybox{Comb} \mybox{Id =}  \mybox{Comb} \mybox{Id x} \mybox{Id x}\linebreak cbow(model, \underline{0.5}*lr, preorder\_line);};
\node(inorder)[inorder,below of=preorder,xshift=+15.0mm,yshift=-3.8mm]{\mybox{Id !} \mybox{Comb} \mybox{Id num} \mybox{Abs x} \mybox{Id =} \mybox{Comb}  \mybox{Id x} \mybox{Comb} \mybox{Id x}\linebreak cbow(model, \underline{0.3}*lr, inorder\_line);};
\node(postorder)[postorder,below of=inorder,xshift=-1.4mm,yshift=-6.5mm]{\mybox{Id !} \mybox{Id num} \mybox{Id =}  \mybox{Id x} \mybox{Id x} \mybox{Comb}  \mybox{Comb} \mybox{Abs x} \mybox{Comb}\linebreak cbow(model, \underline{0.2}*lr, postorder\_line);};
\end{tikzpicture}
\end{center}
\caption{Preorder, inorder, and postorder traversals of a simple theorem $\forall x:\text{num}. \left(x = x\right)$, with weights 0.5, 0.3, 0.2, respectively. Example illustrated by calling the CBOW method of \texttt{fasttext}, where \texttt{lr} is the learning rate and the third argument contains the token sequence above it. Inside the CBOW method, for each token, a randomized span of words surrounding that token is obtained to compute the hidden vector.}
\label{fig:traverse}
\end{figure}

Both DeepWalk and Node2Vec directly use the Word2Vec implementation of the Gensim~\cite{gensim} topic modeling library.
For ease of future integration into proof assistants we pick a dedicated Word2Vec implementation \texttt{fasttext}~\cite{joulin-etal-2017-bag} as our base platform.
To make our customization less intrusive we add a custom tree traversal module to the codebase of \texttt{fasttext}, also called the \texttt{tt} module (Fig.~\ref{fig:arch}).
The \texttt{tt} module parses the terms in the \texttt{tt} format and builds corresponding trees in the memory of JEFL.

\subsection{The SGD Updates of Word2Vec}\label{ss:w2v}

It remains to discuss the core Word2Vec algorithm, which is divided into two aspects: 1.~what is the probability model for Word2Vec training and 2.~how the loss function is computed.
In the former, there are the \textit{continuous bag-of-words} model (CBOW) and the \textit{skip-gram} model.
They appear at the step of the training loop outside stochastic gradient descent (SGD) updates and determine how data samples are used.
In the latter, there are the \textit{softmax} loss, the \textit{hierarchical softmax} loss, and the \textit{negative sampling} loss.
They compute the loss function, at the same time determine the gradients and update the input and output matrices.
The two training models are compatible with the three loss functions, so there are in total six variations of the Word2Vec algorithm\footnote{As to writing of this paper, one more loss function (the \texttt{one-vs-all}, or the \texttt{ova} loss) has been added to the latest version of \texttt{fasttext}, making in total eight variations.}.

As the full Word2Vec algorithm is extensive, we briefly describe the difference between skip-gram and CBOW using the simplest softmax case.
We skip detailed derivations and remind the reader of the abundance of study materials of Word2Vec on the internet\footnote{The first author finds this note \href{https://github.com/renpengcheng-github/nlp/tree/master/3.word2vec}{https://github.com/renpengcheng-github/nlp/tree/master/3.word2vec} (in Chinese) particularly helpful in understanding Word2Vec.}.

Let $\mathcal{C}$ be the training corpus, $V$ be the size of the dictionary of $\mathcal{C}$, and $D$ be the dimension of a word vector.
Denote $M\in\mathbb{R}^{V\times D}$ as the \textit{input matrix} which we use to store all the word vectors.
Denote $N\in\mathbb{R}^{V\times D}$ as the \textit{output matrix} which we use to store customized data items depending on the loss function.
Let $w \in \left\{ 1, 2, \ldots, V \right\}$ be a word, or more precisely, the index of an actual word in the dictionary.
We denote $M_w$ as the $w$-th row of the input matrix $M$.
Similarly we denote $N_u$ as the $u$-th row of the output matrix $N$, given a word $u \in \left\{ 1, 2, \ldots, V \right\}$.
Both $M_w$ and $N_u$ are $D$-dimensional row vectors.
For each word $w$, denote $\text{context}(w)$ as a randomized span of words surrounding $w$.
Let $\eta > 0$ be the learning rate.

From the probability modeling point of view, CBOW amounts to maximizing the log-likelihood of the form
\begin{align*}
\mathcal{L}
&= \log\prod_{w \in \mathcal{C}}P\left(w|\text{context}(w)\right)
= \sum_{w \in \mathcal{C}} \log \text{softmax} (Nh^{T})_w,
\end{align*}
where
\begin{align*}
h = \frac{1}{|\text{context}(w)|} \sum_{u \in \text{context}(w)} M_u
\end{align*}
is the hidden vector. The SGD updates are computed by taking \textit{increments} of the gradients of the objective (as we want to \textit{maximize} the log-likelihood)~\footnote{We use the term stochastic gradient \textit{descent} here for convention though we are in fact doing stochastic gradient \textit{ascent}.}
\begin{align*}
N_u
&:= N_u + \eta\left(\delta_{uw} - \text{softmax}(Nh^T)_u\right)h
& \text{for } u \in \{1,\ldots,V\}
& \\
M_u
&:= M_u + \frac{\eta}{|\text{context}(w)|}\sum_{v = 1}^{V}\left(\delta_{vw} - \text{softmax}(Nh^T)_v\right)N_v
& \text{for } u \in \text{context}(w)
\end{align*}
The skip-gram model amounts to maximizing the log-likelihood of the following form
\begin{align*}
\mathcal{L}
&= \log\prod_{w \in \mathcal{C}}P\left(\text{context}(w)|w\right)
= \log\prod_{w \in \mathcal{C}}\prod_{u\in\text{context}(w)}p(u|v) \\
&= \sum_{w \in \mathcal{C}}\sum_{u\in\text{context}(w)}\log\text{softmax}(Nh^T)_u
\end{align*}
where
\[
h = M_w
\]
is a $D$-dimensional row vector. For each $u\in\text{context}(w)$, the SGD updates are
\begin{align*}
N_{\widetilde{w}}
&:= N_{\widetilde{w}} + \eta\left(\delta_{\widetilde{w}u} - \text{softmax}(Nh^T)_{\widetilde{w}}\right)M_w
& \text{for } \widetilde{w}\in\left\{1,\ldots,V\right\} \\
& \\
M_w
&:= M_w + \eta\sum_{v = 1}^{V}\left(\delta_{vu} - \text{softmax}(Nh^T)_v\right)N_v.
\end{align*}

\begin{algorithm}
\begin{algorithmic}[1]
\For {$w\in\mathcal{C}$}
\State Get sample $(w, \text{context}(w))$. \Comment See Section~\ref{ss:adapt}
\If {CBOW}
\State {$h := \mathbf{0}$}
\For {$v \in \text{context}(w)$}
\State {$h := h + M_v$}
\EndFor
\State {$h := h / |\text{context}(w)|$} \Comment 1. Get hidden vector (cbow)
\State $g := \mathbf{0}$
\For {$u \in \{1,\ldots,V\}$} \Comment Room for speedup
\State $s := \text{softmax}\left(Nh^T\right)_u$
\State $\alpha := \eta\left(\delta_{uw} - s\right)$
\State $g := g + \alpha N_u$ \Comment 2. Accumulate gradient (cbow)
\State $N_u := N_u + \alpha h$ \Comment 3. Update output matrix (cbow)
\EndFor
\State $g := g / |\text{context}(w)|$
\For {$u \in \text{context}(w)$}
\State $M_u := M_u + g$ \Comment 4. Update input rows (cbow)
\EndFor
\Else \Comment Skip-gram
\State {$h := M_w$} \Comment 1. Get hidden vector (skipgram)
\For {$u\in\text{context}(w)$}
\State $g := \mathbf{0}$
\For {$\widetilde{w}\in\left\{1,\ldots,V\right\}$} \Comment Room for speedup
\State $s := \text{softmax}\left(Nh^T\right)_{\widetilde{w}}$
\State $\alpha := \eta\left(\delta_{\widetilde{w}u} - s\right)$
\State $g := g + \alpha N_{\widetilde{w}}$ \Comment 2. Accumulate gradient (skipgram)
\State $N_{\widetilde{w}} := N_{\widetilde{w}} + \alpha h$ \Comment 3. Update output matrix (skipgram)
\EndFor
\State $M_w := M_w + g$ \Comment 4. Update input rows (skipgram)
\EndFor
\EndIf
\EndFor
\end{algorithmic}
\caption{Full algorithm for CBOW and skip-gram with softmax loss}
\label{alg:sgd-softmax}
\end{algorithm}

\subsection{The \texttt{fasttext} Implementation of Word2Vec}\label{ss:fasttext}

The full SGD update algorithm is shown in Algorithm~\ref{alg:sgd-softmax}.
Notice that, for both CBOW and skip-gram, in each round of model updates there are essentially four identical steps: 1. obtain hidden vector, 2. accumulate gradient, 3. update rows of the output matrix, and 4. update rows of the input matrix.
This four-step abstraction is general not only for softmax but also for hierarchical softmax and negative sampling, which are specifically designed to speed up the calculation of the inner loop in line 10, 24 of Algorithm~\ref{alg:sgd-softmax}.

The architecture of \texttt{fasttext} was inspired by this four-step abstraction.
Since its initial development in 2016, lots of advanced functionalities have been added on top of the Word2Vec algorithm, including model quantization, autotuning, python binding, etc.
This makes the codebase large and many of those functionalities are irrelevant to our research.
Therefore we use an earlier commit in late 2016 as our base\footnote{\texttt{c62abb89396a94520f009f9095874953735e0d75}}.
In this commit, all six variations of Word2Vec have been implemented and very few advanced functionalities are added.
Two of them worth mentioning are: 1.~subsampling of most frequent words, and 2.~subword information enrichment trick.
The first is an extension of the Word2Vec algorithm in~\cite{word2vec-b} to filter out disproportionally frequent words in the training corpus.
This function is disabled since it is obvious in our dataset that the most frequent tokens are always \texttt{Comb}, \texttt{Id}, and \texttt{Abs}, respectively, and they have to be included to allow for correct parsing of \texttt{tt} items.
The second is a feature in the \texttt{fasttext} implementation~\cite{joulin-etal-2017-bag} which breaks a word token into segments of character-level n-gram tokens.
This is also disabled since constants in our embedding task (e.g. \texttt{'const/arith/PRE'}) constitute a unique and separate entity, and enabling this feature would normally increase the size of a training model by more than a hundred times.
The original \texttt{src} directory of this commit contains 2054 lines of C++ code written in C++11.

\section{Experimenting with JEFL}\label{s:exp}


The core algorithm part of JEFL consists of 8 modules of the original \texttt{fasttext} plus a custom module for term parsing and traversal (Fig.~\ref{fig:arch} right).
We find that the best way for customization is to add to the \texttt{args} module a new flag (\texttt{isTT}) to denote whether our training corpus is a list of \texttt{tt} items or plain texts.
The normal process flow is not interrupted if this flag is \texttt{false}, so JEFL can also train on plain text.
If \texttt{isTT} is \texttt{true}, then
subsampling
is suppressed
when reading in the input files.
This allows all tokens of \texttt{tt} terms to be read so that term parsing can be done correctly.
The \texttt{tt} items are read, parsed
and the parsed terms can be reconstructed as trees in the C++ side.
Helper functions are then called to traverse a tree in different orders, look up the index values of its constants from a dictionary, and call \texttt{fasttext}'s original CBOW or skip-gram method for SGD updates (Fig.~\ref{fig:traverse}).

We report our initial round of experiments with
this platform and
two formal proof
libraries, HOL4 and HOL Light, testing the performance of different hyper-parameter combinations.
There are in total 18723 and 16874 lines of \texttt{tt} items in HOL4 and HOL Light, respectively.
We concatenate and shuffle the exported theorems from the two libraries, and then write them out as s-expressions~\cite{mccarthy_recursive_1960}.\footnote{This gives a total of 35597 s-expressions for Word2Vec training.}
We evaluate the performance by comparing against the 1000 highest-scoring constant pairs, that have been manually checked by Gauthier in his work, and considered here as a baseline.
%
%

\begin{table}[h!]
\centering
\begin{tabular}{p{3cm} p{2cm} p{2cm} p{2cm} p{2cm}} 
\hline
\ & Top-1 Hit & Top-3 Hit & Top-10 Hit & Top-20 Hit \\ [0.5ex]
\hline
Tree-Dump & 51 & 101 & 188 & 261 \\
Leaf-Dump & 21 & 61 & 96 & 144 \\ [1ex] 
\hline
\end{tabular}
\caption{Comparison of theorem export formats. Tree-dump exports the whole tree representation in the given order, while leaf-dump exports only the sequence of data present in the leafs.}
\label{tab:exp1}
\end{table}

\begin{table}[h!]
\centering
\begin{tabular}{p{3cm} p{2cm} p{2cm} p{2cm} p{2cm}} 
\hline
\ & Top-1 Hit & Top-3 Hit & Top-10 Hit & Top-20 Hit  \\ [0.5ex]
\hline
Skip-Gram & 54 & 108 & 264 & 283 \\
CBOW & 51 & 101 & 188 & 261 \\ [1ex] 
\hline
\end{tabular}
\caption{Comparison of models skip-gram and continuous bag of words.}
\label{tab:exp2}
\end{table}

\begin{table}[h!]
\centering
\begin{tabular}{p{3cm} p{2cm} p{2cm} p{2cm} p{2cm}}
\hline
\ & Top-1 Hit & Top-3 Hit & Top-10 Hit & Top-20 Hit \\ [0.5ex]
\hline
Hierarchical Softmax & 78 & 161 & 304 & 419 \\
Negative Sampling & 51 & 101 & 188 & 261 \\ [1ex]
\hline
\end{tabular}
\caption{Comparison of sampling hierarchical softmax vs. negative sampling}
\label{tab:exp3}
\end{table}

\begin{table}[h!]
\centering
\begin{tabular}{p{3cm} p{2cm} p{2cm} p{2cm} p{2cm}} 
\hline
(pre-,in-,post-order) & Top-1 Hit & Top-3 Hit & Top-10 Hit & Top-20 Hit \\ [0.5ex]
\hline
(0.33,0.33,0.33) & 51 & 101 & 188 & 261 \\
(0.5,0.3,0.2) & 58 & 103 & 205 & 267 \\
(1,0,0) & 53 & 110 & 204 & 256 \\
(0.5,0.5,0) & 49 & 113 & 207 & 276 \\
(0,0.5,0.5) & 57 & 106 & 203 & 268 \\ [1ex]
\hline
\end{tabular}
\caption{Combination of the effect of weights given to the different traversals. The table shows the weights given to pre-order, in-order, and post-order respectively, together with their effects on finding same constants across libraries.}
\label{tab:exp4}
\end{table}

We present four sets of experiments for the initial comparison.
We measure JEFL's performance against the Gauthier baseline by using the ``Top-$N$ Hit'' metric, which means the inclusion of the correct answer from the closest $N$ neighbors of the target constant.
By default, we use leaf-dump (sequences of data present in the leafs), CBOW, negative sampling, and equal (0.33,0.33,0.33) weights.
For other key parameters of \texttt{fasttext}, we set vector dimension as 100, learning rate 0.05, random uniform context window size 1 to 10, training epoch 5 (for most experiments we see little training progress after epoch 5, so for a fair evaluation we stick with 5 epochs for all evaluations), 5 negative samples for negative sampling loss, and 4 training threads.

Experiment 1 (Table \ref{tab:exp1}) tests the difference between tree-dump (dumping s-expression) vs. leaf-dump (dumping leaves as token sequences).
This experiment is the first one to test that our customization blends with the normal process flow of \texttt{fasttext}.
We see that tree traversal gives better hit rates than just use the leaves as the former uses more information in training.

Experiment 2 (Table \ref{tab:exp2}) tests the difference between skip-gram and CBOW.
We see that skip-gram performs better than CBOW in all hit rates.
However, as noted in Section~\ref{ss:w2v}, skip-gram takes longer time to train (this depends on the size of the contexts, and in our experiments skip-gram takes about five times longer).
For ease of experiment, we fall back to use CBOW as our default.

Experiment 3 (Table \ref{tab:exp3}) shows a clear advantage of hierarchical softmax over negative sampling.
We see a 60\% increase in all the hit rates.
We speculate that this improvement is due to the fact, that the Huffman tree computation in hierarchical softmax might put an advantage in mining patterns in tree-like data structures.

Experiment 4 (Table \ref{tab:exp4}) explores different combinations of weights in tree-traversal.
They all outperform leaf-dump, however, none of the combinations performs significantly better than others.
We plan to explore other forms of traversal such as random walks to see further results.

\section{Comparison with Iterative Pattern-Matching}\label{s:comparison}

In this section, we shortly recall the iterative pattern-matching
algorithm developed by Gauthier and Kaliszyk~\cite{tgck-jsc19}, and compare it
with the work presented here.
The iterative pattern matching algorithm is based on the observation that once mathematical information in different formal libraries is represented in the same \texttt{tt} format, similar theorems or typing judgements (as terms of \texttt{tt}) tend to have identical term structures.
Accordingly, similar constants (as leaves of terms) tend to locate in corresponding slots of a term (Fig.~\ref{fig:patt}).
To abstract out common term structure, Gauthier invented the notion \textit{pattern of a term}.
The pattern of a term $T$ is created by abstracting out, in a canonical order, all the $T$'s non-logical constants.
Two terms $T_1$ and $T_2$ sharing the same pattern form a \textit{matching pair of terms}.
Corresponding slots of a matching pair of terms \textit{induce} a collection of \textit{matching pairs of constants}.

\tikzstyle{term}=[text width=1em, text centered, font=\ttfamily\scriptsize]
\tikzstyle{slot}=[draw,fill=red!20,rounded rectangle,dashed,text centered,font=\ttfamily\scriptsize]
\tikzstyle{toplabel}=[xshift=5mm, yshift=6mm,font=\ttfamily\scriptsize]
\tikzstyle{botlabel}=[xshift=5mm, yshift=-10mm,font=\ttfamily\scriptsize]

\begin{figure}[h!]
\begin{tikzpicture}[level distance=7mm, sibling distance=10mm]
\node(x01)[term]{Comb}
 child { node(x02)[term]{Id(!)} }
 child {
  node(x03)[term]{Abs(x)}
  child { node(x05)[term]{Id(num)} }
  child {
   node(x06)[term]{Comb}
   child { node(x07)[term]{Id(=)} }
   child {
    node(x08)[term]{Comb}
    child {
     node(x09)[term]{Comb}
     child { node(x11)[term]{Id(+)} }
     child {
      node(x12)[term]{Comb}
      child { node(x13)[term]{Id(x)} }
      child { node(x14)[term]{Id(0)} }
     }
    }
    child { node(x10)[term]{Id(x)} }
   }
  }
 };
\node(y01)[term,right of=x01,xshift=25mm]{Comb}
 child { node(y02)[term]{Id(!)} }
 child {
  node(y03)[term]{Abs(x)}
  child { node(y05)[slot]{Id($v_1$)} }
  child {
   node(y06)[term]{Comb}
   child { node(y07)[term]{Id(=)} }
   child {
    node(y08)[term]{Comb}
    child {
     node(y09)[term]{Comb}
     child { node(y11)[slot]{Id($v_2$)} }
     child {
      node(y12)[term]{Comb}
      child { node(y13)[term]{Id(x)} }
      child[missing]
      child { node(y14)[slot]{Id($v_3$)} }
     }
    }
    child { node(y10)[term]{Id(x)} }
   }
  }
 };
\node(z01)[term,right of=y01,xshift=25mm]{Comb}
 child { node(z02)[term]{Id(!)} }
 child {
  node(z03)[term]{Abs(x)}
  child { node(z05)[term]{Id(real)} }
  child {
   node(z06)[term]{Comb}
   child { node(z07)[term]{Id(=)} }
   child {
    node(z08)[term]{Comb}
    child {
     node(z09)[term]{Comb}
     child { node(z11)[term]{Id($\times$)} }
     child {
      node(z12)[term]{Comb}
      child { node(z13)[term]{Id(x)} }
      child { node(z14)[term]{Id(1)} }
     }
    }
    child { node(z10)[term]{Id(x)} }
   }
  }
 };
\path[<->,>=stealth']
  (x05) edge[bend left] node [right] {} (y05)
  (y05) edge[bend left] node [right] {} (z05)
  (x11) edge[bend left] node [right] {} (y11)
  (y11) edge[bend left] node [right] {} (z11)
  (x14) edge[bend right] node [right] {} (y14)
  (y14) edge[bend right] node [right] {} (z14);
\node at([toplabel]x01) {$T_1\quad\forall x : \mbox{num}.\ (\ x + 0 = x\ )$};
\node at([toplabel]z01) {$T_2\quad\forall x : \mbox{real}.\ (\ x \times 1 = x\ )$};
\node at([botlabel]y13) {$P\quad\lambda v_1\lambda v_2\lambda v_3.\  (\ \forall x : v_1.\  (\ x = x \ v_2 \ v_3\ ))$};
\end{tikzpicture}
\caption{$T_1$ and $T_2$ form a matching pair of terms with pattern $P$. Three matching pairs of constants can be induced from this pattern. We treat equality \texttt{=} and universal quantification \texttt{!} as logical constants. Bound variables are assumed to be normalized.}
\label{fig:patt}
\end{figure}
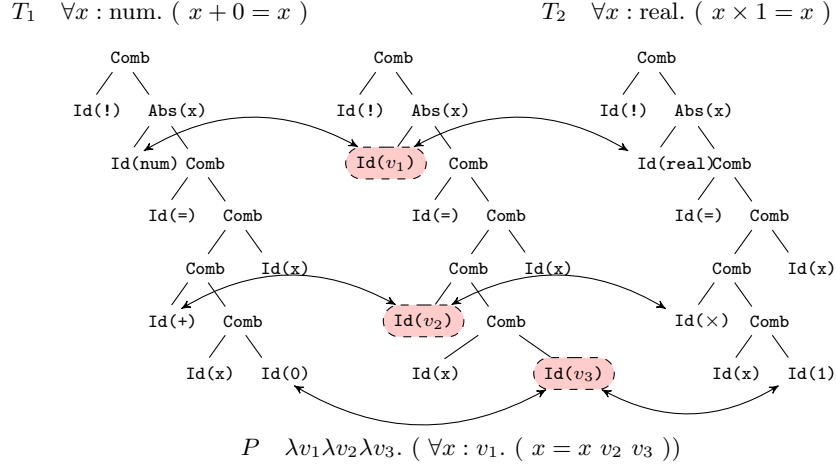

Given two formal libraries $L_1$ and $L_2$.
Let $\{t_i\}_{1\le i\le m}$ be the collection of all matching pairs of terms, with $t_i = (t_{i1}, t_{i2})$, $t_{i1}\in L_1$, and $t_{i2}\in L_2$.
Let $\{c_j\}_{1\le j\le n}$ be the collection of all matching pairs of constants, with $c_j = (c_{j1}, c_{j2})$.
Let $g(x) = x / (x + 1): \mathbb{R}^{+}\rightarrow [0,1)$ be a strictly increasing \textit{normalization function}.
Define an \textit{indicator function} $\delta(c_j,t_i)$ and set $\delta(c_j,t_i) = 1$ if constant pair $c_j$ can be induced by term pair $t_i$ and $0$ otherwise.
Similarity scores between pairs of terms and pairs of constants can be calculated using the following recurrence relations
\begin{equation}
\begin{cases}
\text{score}^0_c\left(c_j\right) = 1, & j = 1,\ldots,n.\\
\text{score}^{T}_t\left(t_i\right) = w_t\left(t_i\right)\sum_{l=1}^n \delta(c_l, t_i)\, \text{score}^{T-1}_c(c_l), & i = 1,\ldots,m.\\
\text{score}^{T}_t\left(c_j\right) = g\left(w_c\left(c_j\right)\sum_{k=1}^m\delta(c_j, t_k)\, \text{score}^{T}_t(t_k)\right), & j = 1,\ldots,n.\label{eq:score}
\end{cases}
\end{equation}
where $T = 0,1,2,\ldots$ is the iteration step and the weighting functions for terms $w_t\left(t_i\right)$ and constants $w_c\left(c_j\right)$ are determined using heuristics
\begin{equation}
\begin{cases}
w_t\left(t_i\right) = \frac{1}{\ln\left(2 + p(t_i)\right)}\frac{1}{\ln\left(2 + q(t_i)\right)}, & i = 1,\ldots,m.\\
w_c\left(c_j\right) = \frac{1}{\ln\left(2 + r(c_{j1})\times r(c_{j2})\right)}, & j = 1,\ldots,n.\\
p\left(t_i\right) = \#\{\text{\small term pairs sharing the same pattern as } t_i\},& i = 1,\ldots,m.\\
q\left(t_i\right) = \#\{\text{\small constant pairs induced by } t_i\},& i = 1,\ldots,m.\\
r\left(c_{jd}\right) = \#\{\text{\small terms containing } c_{jd}\},\quad\quad\quad\quad\quad d = 1\text{ or }2,& j = 1,\ldots,n.
\label{eq:heuristic}
\end{cases}
\end{equation}
By rewriting equation~(\ref{eq:score}) with respect to $\text{score}^{T}_t\left(c_j\right)$ and using properties of $g$, $\delta$, $w_t$, and $w_c$, Gauthier proved the convergence of this scoring algorithm (using monotone convergence theorem coordinate-wise in $[0,1]^n$)~\cite{tgck-jsc19}.

\subsection{Advantages and Drawbacks of Iterative Pattern Matching}

Gauthier's iterative pattern-matching algorithm is cleverly designed.
Intuitively, the existence of a pattern already indicates a strong correlation among term pairs, and the existence of constant pairs at corresponding slots of a pattern already indicates a strong correlation among those constant pairs;
The indicator function $\delta$ transports ``similarity awards'' among those pairs, while the weighting functions $w_t$ and $w_c$ penalize frequently occurring patterns.
Above all, the normalization function $g$ ensures the validity of the scores and is crucial for convergence of the algorithm.
All these components are intricately combined to make the whole algorithm effective in discovering identical or similar mathematical concepts.

Nevertheless, Gauthier's algorithm possesses some inherent drawbacks.
From the~\textit{explainability} angle, the balance between the heuristics ($w_t$, $w_c$ and their components $p$, $q$, and $r$) in equation~(\ref{eq:heuristic}) and the score accumulation terms ($\sum$ and $\delta$) in equation~(\ref{eq:score}) is, to our mind, hard to explain clearly and difficult to readjust.
The convergence of the algorithm is mostly due to the property of the normalization function $g$ but the link between this convergence and how similarity scores are sorted is weak.
The algorithm works on spaces of similarity scores between matching pairs of terms and constants, so from its beginning, the information on non-matching terms and constants is thrown away, losing the possibility to look at the alignment of different proof assistant libraries holistically.

Comparatively, our approach provides much more flexibility and intuitiveness.
Using distributed representation learning, all the constants are movable points in Euclidean space and similarity between constants are naturally described as cosine similarity between their coordinates.
By using an embedding approach, not only the ``matching'' pairs, but also similarity between all pairs of constants can be retrieved and their computation is cheap.

From the~\textit{customizability} angle, the components of Gauthier's algorithm are so intricately combined that there seems little opportunity to adjust the algorithm further.
In the implementations of both~\cite{tgck-cicm14} and~\cite{tgck-jsc19}, most of the extra work is on preprocessing terms using combinations of rewriting rules to create a varying set of patterns.
These rewriting rules include,~\textit{e.g.} rewriting to conjunctive normal forms, reordering commutative/associative connectives, substituting subterms with definitions, as well as exposing various levels of typing information.
All these customizations are only allowed in the preprocessing phase and limited to only employing rewriting rules.
Some of the typing exposure rules require specific knowledge of representing a library in the \texttt{tt} format.
As these rewriting-based customizations have already been thoroughly investigated in~\cite{tgck-jsc19}, it seems to us that further investigation along this line is destined to a diminishing return.

Comparatively, customization of JEFL can be done at different phases of the full training algorithm, such as the data generation phase, term traversal phase, model update phase, etc.
This multi-level customization can create combinatorially much more room for parameter-tuning and experimentation.
Moreover, except data generation, all other customizations that are within the algorithm can be more uniformly done.
Specifically, data fields and command parsing can be added to the \texttt{args} module, and then used at desired places of the data flow.

From the~\textit{online-servability} angle, despite the fact that Gauthier's algorithm is overall fast and effective on small data, it is still quadratic on the size of the input libraries, since it needs to enumerate all pairs of terms to find patterns.
Being a batch program without a separated and instantaneous evaluation phase, it is not tempting to integrate the whole algorithm into an actual interactive proof assistant.
Moreover, even if it were to be used as an online recommender, the only information we can retrieve would be limited to only the matching pairs.

In JEFL, despite more computations are involved, training is linear with respect to the size of the corpus.
This makes JEFL more suitable for obtaining similarity measurements on a large training corpus.
JEFL provides a clear separation between a training phase and an evaluation phase.
The evaluation phase is instant once the model is loaded.
This allows JEFL to have more potential to be integrated as a service.
In the version of the \texttt{fasttext} commit used in JEFL, if subword information is disabled, the size of most of the models dumped after training are less than 5MB.
This is a negligible size comparing to the size of a modern proof assistant.

\section{Conclusion}\label{s:concl}

In this paper, we identify the need for commonality discovery among formal libraries.
We introduce our data pipeline, especially its preceding works, and elaborate our internal \texttt{tt} formalism.
Methodologically, we describe the architecture of JEFL and
make a series of first experiments to test the efficacy of our experiment platform and provide
a 
high-level comparative analysis with the iterative pattern matching algorithm.

\subsection{Limitations and Future Work}

There are a lot of future possibilities in our JEFL platform.
Continuing in the current line of development, we still need to experiment on the other four libraries and additionally explore similarity discovery of not only constants but also terms.
We could also explore the effect of vector initialization in our discovery algorithm.
To go deeper we could implement custom ``dragging'' and ``repelling'' steps using geometric manipulation and intersperse these custom steps with SGD updates.
We have focused on one pair of libraries, which could be extended to an embedding of multiple libraries combined.
This would provide further experiment opportunities. 
We also plan to use the newly discovered samples from JEFL to do tasks such as conjecturing \cite{tgckju-cicm16}, cross-browsing \cite{dmtgckmkfr-cicm17}, and stronger learning for hammers \cite{tgck-lpar15}.
Last but not least, we hope there could be use cases to integrate our pipeline into an actual proof assistant and see improved formalization productivity.

\section*{Acknowledgements}

We are largely indebted to Thibault Gauthier for his work on alignments and the various data exports that we re-use.
We thank Josef Urban for the Mizar export and his invitation to Prague to discuss research.
We also thank Tom{\'{a}}{\v{s}} Mikolov for valuable insights for the current work.
This work was supported by the ERC grant no. 714034 \textit{SMART} and by the \textit{University of Innsbruck PhD scholarship}.

\bibliographystyle{splncs04}
\bibliography{paper}

\end{document}